\documentclass[pra,twocolumn,showpacs]{revtex4}
\usepackage{amsmath}
\usepackage{amssymb}
\usepackage{float}
\usepackage{bm,times}
\usepackage{graphicx}
\newcommand{\be}{\begin{equation}}
\newcommand{\ee}{\end{equation}}
\newcommand{\bea}{\begin{eqnarray}}
\newcommand{\eea}{\end{eqnarray}}
\newcommand{\ek}{\epsilon_{\mathbf{k}}}

\newcommand{\Ek}{E_{\mathbf{k}}}

\newcommand{\uk}{u_{\mathbf{k}}}
\newcommand{\vk}{v_{\mathbf{k}}}

\begin{document}

\title{Finite Temperature Momentum Distribution of a Trapped Fermi Gas}

\date{\today}

\author{Qijin Chen,$^1$ C. A. Regal,$^2$ D. S. Jin,$^2$ and K.
  Levin$^1$}

\affiliation{$^1$James Franck Institute and Department of
Physics, University of Chicago, Chicago, Illinois 60637, USA}

\affiliation{$^2$JILA, Quantum Physics Division National
Institute of Standards and Technology and University of Colorado,
and Department of Physics, University of Colorado, Boulder,
Colorado 80309-0440, USA}

  \date{\today}

\begin{abstract}
  We present measurements of the temperature-dependent momentum
  distribution of a trapped Fermi gas consisting of $^{40}$K in the
  BCS-BEC crossover regime.  Accompanying theoretical results based upon a
  simple mean-field ground state are compared to the experimental data.
  Non-monotonic effects associated
  with temperature, $T$,  arise from the competition between thermal broadening
  and a narrowing of the distribution induced by the decrease in
  the excitation gap $\Delta(T)$ with increasing $T$. 
\end{abstract}

\pacs{03.75.Hh, 03.75.Ss, 74.20.-z \hfill{\textsf{cond-mat/0604469}}}

\maketitle

The recent discovery of superfluidity in trapped fermionic gases has
paved the way for arriving at a much deeper understanding of the
phenomena of superfluidity and superconductivity
\cite{Jin3,Grimm,Jin4,Ketterle3,KetterleV,Thomas2,Grimm3,ThermoScience}.
This may ultimately have application in high temperature superconductors
\cite{ourreview,ReviewJLTP}, nuclear, astro- and particle physics. In
these trapped gases one has the ability to tune the strength of the
attractive interaction that leads to pairs of fermions (Cooper pairs),
which then Bose condense.  As the pairing strength is increased a smooth
crossover from BCS behavior to Bose-Einstein condensation (BEC) occurs.
This tunability is a remarkable feature of trapped atomic gases and
arises from a phenomenon known as a Feshbach resonance. In general the
resulting superfluid state is more complex than that of simple BCS
theory. Similarly the normal state is expected to be quite different
from its BCS analogue, since in general pairing takes place at higher
temperatures ($T^*$) than the condensation temperature ($T_c$).

The ultracold gases have, thus, presented us with an opportunity
to investigate in a more complete fashion all aspects of fermionic
superfluidity.
Because of their charge neutrality and because of trap confinement
effects there exists a different set of tools for their
experimental investigation. Among these tools are measurements of
the real space and momentum space distributions of the atomic
fermions \cite{Salomon,Grimm2,Jin6,Hulet5,Ketterle6}. At finite
temperatures these distributions must change significantly in ways
that reflect both pairing and possibly the onset of phase
coherence.

In this paper we study both experimentally and theoretically the
temperature dependence of the momentum distribution of fermionic
atoms within a trapped Fermi gas.  Recent measurements of this
momentum distribution near $T=0$ showed a dramatic broadening as
the pairing strength increased \cite{Jin6}, as is qualitatively
consistent with theoretical calculations \cite{Viverit,Jin6}.  In
the present paper we extend these measurements by varying the
temperature from below to well above the theoretically predicted
values of $T_c$; this allows us to probe the momentum distribution
in the normal phase as well as the superfluid phase.

In general, in a $T=0$ fermionic system the physical phenomenon
that controls the momentum distribution is pairing. In the BCS
theory the momentum distribution of a homogeneous sample shows a
slight smearing of the Fermi surface due to pairing.  This effect
is very small and associated with the gap parameter $\Delta$.  In
the BCS-BEC crossover, however, one greatly increases the
interaction strength that leads to pairing.  It is conventional to
parameterize the state within this crossover in terms of the
dimensionless product $k_F^0 a $, where $a$ is the two-body
$s$-wave scattering length between atoms and $k_F^0$ is the Fermi
wavevector in the noninteracting limit. While in the BCS limit
($1/k_F^0 a \rightarrow - \infty$) it is difficult to observe the
slight change in the momentum distribution, as the interaction
strength increases this broadening grows.  Indeed, at unitarity
($1/k_F^0 a = 0$) the broadening is no longer small, but instead
comparable to the Fermi energy. Finally, in the BEC limit ($
1/k_F^0 a \rightarrow \infty$) the atomic momentum distribution
becomes extremely broad and corresponds to the square of the
molecular wavefunction in momentum space.

At finite $T$ one expects a similar broadening of the momentum
distribution as the interaction is increased.  However, the extent
of this broadening is determined by the temperature dependence of
the excitation gap $\Delta(T)$; it is maximum at $T=0$ and should
disappear as $\Delta(T)$ goes to zero around $T^*$, where all the
fermion pairs are broken.  This effect of pairing on the
distribution will occur in addition to the usual thermal
broadening of the momentum profile that occurs in a Fermi gas.
While the thermal broadening increases with $T$, the pairing
induced broadening decreases with $T$.  As we will show below,
these two competing effects lead to a non-trivial $T$ dependence
in the Fermi gas momentum distribution.

While the discussion thus far has centered on the expectation for
a homogeneous Fermi gas, the experiments presented here take place
in a harmonic trapping potential.  In such a potential the local
Fermi wavevector varies with position $r$,
so that the integrated momentum distribution for a noninteracting Fermi
gas in a trap is significantly different from that for a homogeneous
case \cite{Jin6}.  Thus, in addition to the local broadening due to
pairing, the trapping potential compresses the density profile and
thereby enlarges the overall momentum distribution.

To perform our experiments, we follow the techniques described in
Refs. \cite{Jin2,Jin} to create an ultracold $^{40}$K gas in the
BCS-BEC crossover.   Briefly, starting with a cold gas we slowly
ramp a magnetic field to approach a wide Feshbach resonance at
$202.1\pm 0.1$~G \cite{Jin4}.  The rate of the magnetic field
ramp, (6.5 ms/G)$^{-1}$, is slow enough to ensure adiabaticity
with respect to many-body time scales in the system.  We then
probe the momentum distribution of atoms in this final state using
the experimental sequence developed in Ref.~\cite{Jin6}.  In this
sequence the atom gas must expand freely without any interatomic
interactions; this we achieve through a fast magnetic-field ramp
(here $\approx 8$  $\mu$s/G) to $a\approx 0$ \cite{Salomon}.  This
is followed by the standard technique of time-of-flight expansion
and optical absorption imaging.

To obtain the data in this paper we repeat the process above for
gases at a variety of interaction strengths and at a variety of
temperatures.  To vary the temperature of our initial Fermi gas we
increase the depth of our optical dipole trap after evaporative
cooling and heat the gas by modulating the optical trap strength.
To characterize the temperature of the gas we measure the
temperature compared to the Fermi temperature in the
non-interacting limit, $(T/T_F)^0$.  The physical temperature at a
given $k_F^0 a$ can then be extracted using the theoretical
thermometry described in Ref. \cite{ChenThermo}; this thermometry
is based upon entropy conservation in the adiabatic magnetic field
ramp.

\begin{figure}
\centerline{\includegraphics[clip,width=3.4in]{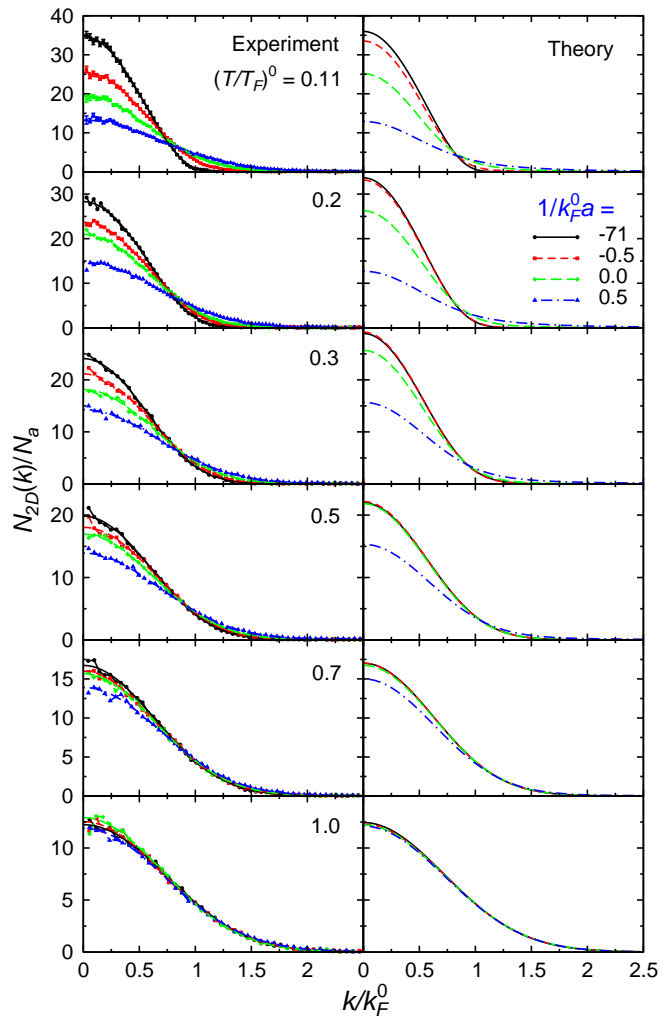}}
\caption{(color online) Evolution of the momentum distribution
  $N_{2D}(k)$ with the interaction strength $1/k_F^0a$ from
  noninteracting to near-BEC.  The two columns compare experiment (left)
  and theory (right). Different rows correspond to different values of
  $(T/T_F)^0$.  The lines in the experimental plots correspond to a fit
  to Eq.~\ref{eq:OD}.  The value of $k_F^0$ used to create the
  experimental plots was determined through experimental measurements of
  the particle number and trap strength, which lead to systematic
  uncertainties in $k_F^0$ of up to 10$\%$ for these data.  While the
  raw data are shown here, the normalization applied in Fig.~\ref{fig:3}
  removes this systematic error.}
\label{fig:1}
\end{figure}

In Fig. 1, we plot experimental azimuthally averaged momentum
distributions as a function of $k/k_F^0$. From top to bottom, each panel
corresponds to a fixed temperature $(T/T_F)^0 = 0.11$, 0.2, 0.3, 0.5,
0.7, and 1.0, with variable $1/k_F^0a$.  The scattering strength
$1/k_F^0a=-71$, -0.5, 0.0, and 0.5 represent the noninteracting Fermi
gas, near-BCS, unitary, and near-BEC cases, respectively.  For each set
of measured distributions in Fig.~1 at a particular $(T/T_F)^0$,
parameters such as trap strength and initial gas density are fixed.
However, these parameters vary among the different sets at constant
$(T/T_F)^0$ (panels in Fig.~\ref{fig:1}). For example, for the data at
$(T/T_F)^0=0.11$ the peak density, for atoms in one spin state, in the
weakly interacting regime is $n_{pk}^0 = 1.4 \times 10^{13}$ cm$^{-3}$
and $E_F^0=0.56$ $\mu$K. For the hottest data $n_{pk}^0$ decreases to $6
\times 10^{12}$ cm$^{-3}$ and $E_F^0=0.79$ $\mu$K.

On top of the data points in Fig.~1 we show a fit to an empirical
function, which we utilize both for normalization and for later
data analysis. We apply the following two-dimensional (2D) surface
fit to the measured optical depth (OD) of the expanded Fermi gas.
\begin{equation}
OD(x,y) = OD_{pk}\, g_2 \Big(-\zeta e^{-\frac{x^2}{2\sigma_x^2}
    -\frac{y^2}{2\sigma_y^2}} \Big)/g_2 (-\zeta),
\label{eq:OD}
\end{equation}
Here $g_n(x) = -\frac{1}{\Gamma(n)} \int_0^\infty \mathrm{d}
\epsilon \frac{\epsilon^{n-1}} {1- e^\epsilon/x} =
\sum_{l=1}^\infty \frac{x^l}{l^n}$ is the poly-logarithmic
function, $\zeta$ is the fugacity, $\sigma^2_{x,y}$ are
proportional to the Fermi gas temperature and related to the
expansion time $t$.  This form can be derived microscopically in
the limit of weak interactions, where $\zeta \rightarrow
e^{\mu/k_BT}$. While this equation is only physically valid for an
ideal Fermi gas, empirically we find that it fits reasonably well
to data throughout the crossover \cite{Jin6}.  In this way the 2D
momentum distribution is given by $ N_{2D}(k_x,k_y) = A$ $OD
(\hbar k_x t /m, \hbar k_y t /m)$. Here $A$ is a normalization
constant such that $\int \frac{\mathrm{d}^2 k }{(2\pi)^2}
N_{2D}(k_x,k_y) = N_a$, where $N_a$ is the total number of atoms.

The experimental results of Fig. 1 show that, at $(T/T_F)^0 = 0.11$, as
the system passes from an ideal Fermi gas to near-BEC the momentum
distribution widens significantly, just as the $T=0$ case shown in Ref.
\cite{Jin6}.  As $(T/T_F)^0$ increases the effect of pairing becomes
less pronounced, and at sufficiently high $T$ the curves coalesce.  Here
the gap $\Delta(T)$ is small compared to $T$, and the system basically
behaves as a classical gas of atoms.

\begin{figure}
\centerline{\includegraphics[clip,width=2.8in]{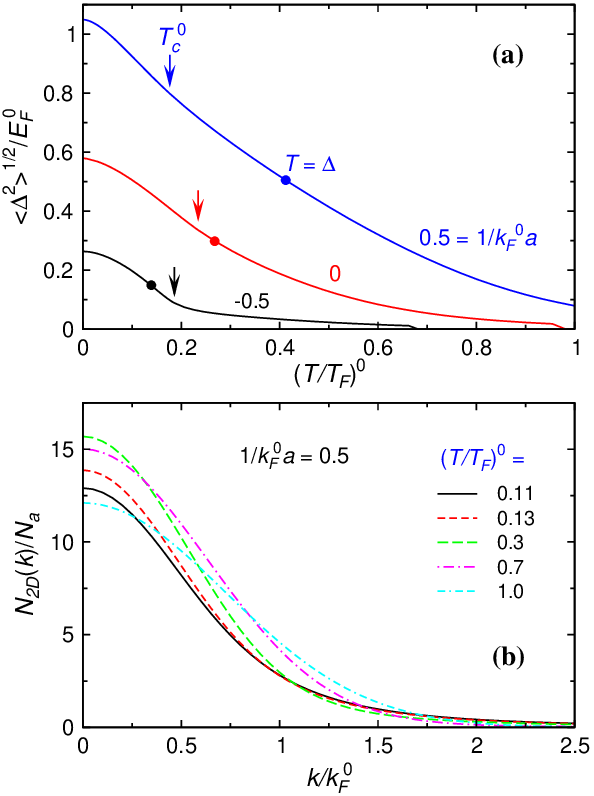}}
\caption{ (color online) (a) Temperature dependence of the average
  excitation gap $\langle \Delta^2\rangle^{1/2}$ at $1/k_F^0a=0.5$ (top
  blue curve), 0 (red) and -0.5 (bottom black curve). The solid circles
  indicate where this effective gap coincides with the temperature. The
  arrows indicate the transition temperatures $T_c^0$.  (b) Theoretical
  momentum distributions $N_{2D}(k)$ at a variety of temperatures for
  $1/k_F^0a=0.5$.}
\label{fig:2}
\end{figure}

Theoretical calculations for the momentum distributions in the crossover
are presented in the right column of Fig. \ref{fig:1}.  The calculations
are based on a generalized mean-field theory \cite{ourreview,JS5}, which
is consistent with the BCS-Leggett ground state \cite{Leggett}.  While
$T=0$ momentum profiles of this state have been discussed in the
literature \cite{Stringari,Jin6}, here we include finite temperature
effects.
An important aspect of this theory is that the fermionic excitation gap
$\Delta(T)$ becomes distinct from the superfluid order parameter
$\Delta_{sc}(T)$ at finite $T$ (except in the strict BCS regime).  This
point is illustrated in Fig. \ref{fig:2}(a). Here the trap-averaged
value of the excitation gap, $\langle \Delta^2\rangle^{1/2}$, is shown
as a function of $(T/T_F)^0$ for the near-BEC ($1/k_F a =$ 0.5), unitary
($1/k_F a = 0$), and near-BCS ($1/k_F a = -0.5$) cases from top to
bottom.  The arrows indicate the calculated value of $T_c^0$.  We find
that $\Delta(T)$ has a gentle onset at high $T$, corresponding to the
pair formation temperature $T^*$, which is generally much higher than
$T_c$.

The momentum and density distribution, along with the chemical potential
$\mu$, the gap parameter, and the order parameter, are self-consistently
determined using the local density approximation.  The local momentum
distribution is given by
\begin{equation}
n_k(r) = 2\, [ \vk^2 (1-f(\Ek)) + \uk^2 f(\Ek)],
\end{equation}
where $\uk^2,\vk^2= [1\pm (\ek-\mu(r))/\Ek(r)]/2 $ are the BCS coherence
factors; $\ek=\hbar^2 k^2/2m$ is the kinetic energy of free fermions,
$\Ek = \sqrt{(\ek-\mu(r))^2 +\Delta^2 }$  the fermionic quasiparticle
dispersion, and $f(x)$ the Fermi-Dirac distribution function. Then
the trap-integrated momentum distribution $N(k)$ and its 2D projection
are given, respectively, by
\begin{equation}
  N(k) = \int\! \mathrm{d}^3 r\, n_k(r) \quad \text{and} \quad N_{2D}(k) = \int
 \! \frac{dk_z}{\!2\pi\,}\,  N(k).
\end{equation}
Note, it is the 2D distribution $N_{2D}(k)$ that is directly
comparable with our experimentally measured optical depth.  It can
be seen that the effects of superfluidity and of pairing more
generally enter only through $\Ek$ which, in turn, depends upon
$\Delta(T)$. Thus the momentum distribution represents, in effect,
the behavior of the fermionic excitation gap, not the superfluid
order parameter $\Delta_{sc}$. Details of the theoretical
formalism can be found in Ref. \cite{ourreview}.

Near $T=0$ it has been shown that the simple mean-field ground state we
use here semi-quantitatively describes the momentum distribution and
effective kinetic energy \cite{Jin6,Chiofalo2005a}.  Thus, this ground
state is an excellent starting point for finite $T$ comparisons between
theory and experiment. Quantitatively, weaknesses of the mean-field
ground state in reproducing the momentum distribution exist, as seen in
Ref. \cite{Chiofalo2005a}. These weaknesses include that, in the BEC
regime, the inter-boson scattering length is overestimated by roughly a
factor of three. However, this does not substantially affect the
momentum distribution of atoms we address, in part because for the
$k_F^0 a$ considered here we confine our attention to the fermionic
regime ($\mu>0$).  (For $1/k_F^0 a = 0.5$, our calculation shows
$\mu(r=0)/E_F^0 = 0.23$ at $T_c$.) Secondly, the theory does not include
the Hartree self energy. We also do not take into account the finite
rate of the magnetic-field ramp above the Feshbach resonance as was done
theoretically in Ref. \cite{Jin6}. This finite ramp rate leads to slight
redistribution of the momentum so that high energy spectral weight is
somewhat suppressed.  This effect is crucial for accurately comparing to
the effective kinetic energy \cite{Jin6,Chiofalo2005a}, but for our ramp
rates and in the fermionic regime we expect the effect on our
distribution comparisons to be small.

We now focus on the comparison between our mean-field calculations and the
experiment shown in Fig. 1. Overall we find semi-quantitative agreement.
However, there are discrepancies that can
be seen in the profiles especially in the middle range of $1/k_F^0a$.
This can be attributed mainly to the neglect of the Hartree term in the
theoretical formalism that was noted above.  If we were to include
the Hartree correlation our calculation of the chemical potential $\mu$
would decrease; this would result in a wider spread between the
noninteracting and interacting profiles in the theoretical portion of
Fig.~\ref{fig:1} and, thus, a better quantitative agreement with
experiment.

\begin{figure}
\centerline{\includegraphics[clip,width=2.8in]{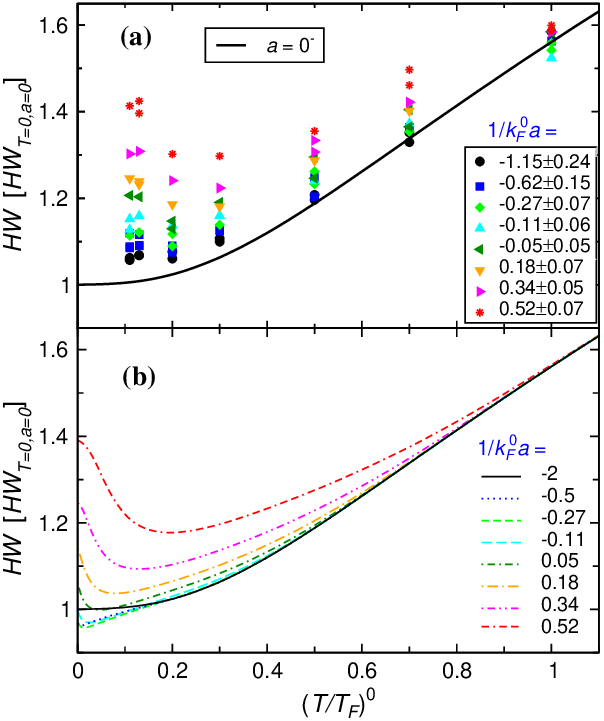}}
\caption{(color online) Comparison of the half width at half height
  ($HW$) as a function of $(T/T_F)^0$ for different values of $1/k_F^0a$
  from noninteracting (bottom curve) to near-BEC (top curve) between (a)
  experiment and (b) theory. The $HW$ in the experimental case is
  normalized to eliminate dependence upon the experimental determination
  of $E_F^0$ (see text); this results in perfect agreement with theory
  in the non-interacting case (black line).}
\label{fig:3}
\end{figure}

We now turn to understanding the $T$ dependence of the
distribution at a fixed value of $1/k_F^0a$. In this analysis one
expects a competition between conventional thermal broadening of
the momentum profiles and narrowing of the profiles due to the
decrease in size of $\Delta(T)$ with increasing temperature.  To
illustrate this quantitatively we turn to Fig. \ref{fig:2}(a). The
solid circles in this figure indicate where the effective gap
coincides with the temperature.  This corresponds roughly to the
temperature at which the broadening as a function of $T$
should display an inflection point.

In Fig. \ref{fig:2}(b) we plot the theoretical results for the
momentum distribution at a fixed $1/k_F^0a=0.5$.  Indeed, there is
a change in the temperature dependence of the curves which occurs
roughly when $\langle\Delta^2\rangle^{1/2} = k_BT$. Below this
temperature the behavior of the distribution tends to be dominated
by that of the excitation gap, and the distribution initially
narrows with increasing $T$. Above this temperature, thermal
effects dominate and the distribution widens with increasing $T$.
We find analogous non-monotonic behavior at the unitary limit.

To test this theoretical prediction of a non-monotonic temperature
dependence of the broadening, we plot the half width at half height
($HW$) of the distributions as a function of $(T/T_F)^0$.  The
experimental $HW$ is determined from a fit of the data to
Eq.~\ref{eq:OD} using the curves in Fig. \ref{fig:1} as well as
additional similar data. For this analysis we have eliminated the
dominant uncertainty associated with the determination of $E_F^0$ (and
hence $k_F^0$) by applying a multiplicative correction factor; this
factor is simply the ratio of the calculated and measured half widths
for the weakly interacting limit.  This is done for each set of data
taken at fixed $(T/T_F)^0$.

Figure \ref{fig:3}(a) shows the experimental results; the different
symbols represent groups of data with an average value of $1/k_F^0a$
indicated by the legend. The black line represents the theoretical
dependence of the $HW$ in the non-interacting ($a=0$) limit.  As
expected for this limit, in the classical regime the width scales
linearly with the temperature; the width then levels out to a finite
value due to Pauli pressure. As the interaction is increased (black
circles to red stars) the decrease of the width with decreasing $T$
becomes less dramatic, and at the lowest $T$ the width actually begins
to increase as a function of $T$.  This temperature non-monotonicity is
particularly apparent on the BEC side of the Feshbach resonance.

The experimental result of Fig. \ref{fig:3}(a) compares favorably with
its theoretical counterpart (Fig. \ref{fig:3}(b)) in that they both
display non-monotonic $T$ dependence. Note that the theoretical $HW$ at
low $T$ and weak interactions is smaller than its experimental
counterpart; in this regime the width drops below the noninteracting
curve due to the absence of the Hartree term and changes in the shape of
the distribution. This aspect of the theory result again accentuates
that, while we have found the simple mean-field theory to be an
excellent starting point for finite temperature studies, future
quantitative studies will require theories that incorporate the Hartree
term.

In summary, we have studied the momentum distribution of trapped
fermionic atoms at finite temperature by a detailed comparison of
theory and experiment. Our results show that there is a
competition between the temperature dependence of the fermionic
excitation gap and thermal broadening, which leads to
non-monotonicities in the temperature dependence of the momentum
profiles. Since temperature is often difficult to determine
experimentally, systematic studies at non-zero $T$ are just
beginning in these ultracold gases. Here, by working with a
theoretical temperature scale (set by the entropy), under
conditions of an adiabatic sweep, we are able to show
semi-quantitative agreement between theory and experiment using a
simple mean-field theory.

We thank Cheng Chin for helpful comments.  This work was supported
by NSF, NASA, and NSF-MRSEC Grant No.~DMR-0213745. C.A.R.
acknowledges support from the Hertz foundation.

\bibliographystyle{apsrev}


\begin{thebibliography}{23}
\expandafter\ifx\csname natexlab\endcsname\relax\def\natexlab#1{#1}\fi
\expandafter\ifx\csname bibnamefont\endcsname\relax
  \def\bibnamefont#1{#1}\fi
\expandafter\ifx\csname bibfnamefont\endcsname\relax
  \def\bibfnamefont#1{#1}\fi
\expandafter\ifx\csname citenamefont\endcsname\relax
  \def\citenamefont#1{#1}\fi
\expandafter\ifx\csname url\endcsname\relax
  \def\url#1{\texttt{#1}}\fi
\expandafter\ifx\csname urlprefix\endcsname\relax\def\urlprefix{URL }\fi
\providecommand{\bibinfo}[2]{#2}
\providecommand{\eprint}[2][]{\url{#2}}

\bibitem[{\citenamefont{Greiner et~al.}(2003)\citenamefont{Greiner, Regal, and
  Jin}}]{Jin3}
\bibinfo{author}{\bibfnamefont{M.}~\bibnamefont{Greiner}},
  \bibinfo{author}{\bibfnamefont{C.~A.} \bibnamefont{Regal}}, \bibnamefont{and}
  \bibinfo{author}{\bibfnamefont{D.~S.} \bibnamefont{Jin}},
  \bibinfo{journal}{Nature} \textbf{\bibinfo{volume}{426}},
  \bibinfo{pages}{537} (\bibinfo{year}{2003}).

\bibitem[{\citenamefont{Jochim et~al.}(2003)}]{Grimm}
\bibinfo{author}{\bibfnamefont{S.}~\bibnamefont{Jochim}} \bibnamefont{et~al.},
  \bibinfo{journal}{Science} \textbf{\bibinfo{volume}{302}},
  \bibinfo{pages}{2101} (\bibinfo{year}{2003}).

\bibitem[{\citenamefont{Regal et~al.}(2004)\citenamefont{Regal, Greiner, and
  Jin}}]{Jin4}
\bibinfo{author}{\bibfnamefont{C.~A.} \bibnamefont{Regal}},
  \bibinfo{author}{\bibfnamefont{M.}~\bibnamefont{Greiner}}, \bibnamefont{and}
  \bibinfo{author}{\bibfnamefont{D.~S.} \bibnamefont{Jin}},
  \bibinfo{journal}{Phys. Rev. Lett.} \textbf{\bibinfo{volume}{92}},
  \bibinfo{pages}{040403} (\bibinfo{year}{2004}).

\bibitem[{\citenamefont{Zwierlein et~al.}(2004)\citenamefont{Zwierlein, Stan,
  Schunck, Raupach, Kerman, and Ketterle}}]{Ketterle3}
\bibinfo{author}{\bibfnamefont{M.~W.} \bibnamefont{Zwierlein}},
  \bibinfo{author}{\bibfnamefont{C.~A.} \bibnamefont{Stan}},
  \bibinfo{author}{\bibfnamefont{C.~H.} \bibnamefont{Schunck}},
  \bibinfo{author}{\bibfnamefont{S.~M.~F.} \bibnamefont{Raupach}},
  \bibinfo{author}{\bibfnamefont{A.~J.} \bibnamefont{Kerman}},
  \bibnamefont{and} \bibinfo{author}{\bibfnamefont{W.}~\bibnamefont{Ketterle}},
  \bibinfo{journal}{Phys. Rev. Lett.} \textbf{\bibinfo{volume}{92}},
  \bibinfo{pages}{120403} (\bibinfo{year}{2004}).

\bibitem[{\citenamefont{Zwierlein et~al.}(2005)\citenamefont{Zwierlein,
  Abo-Shaeer, Schirotzek, and Ketterle}}]{KetterleV}
\bibinfo{author}{\bibfnamefont{M.~W.} \bibnamefont{Zwierlein}},
  \bibinfo{author}{\bibfnamefont{J.~R.} \bibnamefont{Abo-Shaeer}},
  \bibinfo{author}{\bibfnamefont{A.}~\bibnamefont{Schirotzek}},
  \bibnamefont{and} \bibinfo{author}{\bibfnamefont{W.}~\bibnamefont{Ketterle}},
  \bibinfo{journal}{Nature} \textbf{\bibinfo{volume}{435}},
  \bibinfo{pages}{1047} (\bibinfo{year}{2005}).

\bibitem[{\citenamefont{Kinast et~al.}(2004)\citenamefont{Kinast, Hemmer, Gehm,
  Turlapov, and Thomas}}]{Thomas2}
\bibinfo{author}{\bibfnamefont{J.}~\bibnamefont{Kinast}},
  \bibinfo{author}{\bibfnamefont{S.~L.} \bibnamefont{Hemmer}},
  \bibinfo{author}{\bibfnamefont{M.~E.} \bibnamefont{Gehm}},
  \bibinfo{author}{\bibfnamefont{A.}~\bibnamefont{Turlapov}}, \bibnamefont{and}
  \bibinfo{author}{\bibfnamefont{J.~E.} \bibnamefont{Thomas}},
  \bibinfo{journal}{Phys. Rev. Lett.} \textbf{\bibinfo{volume}{92}},
  \bibinfo{pages}{150402} (\bibinfo{year}{2004}).

\bibitem[{\citenamefont{Bartenstein
  et~al.}(2004{\natexlab{a}})\citenamefont{Bartenstein, Altmeyer, Riedl,
  Jochim, Chin, Denschlag, and Grimm}}]{Grimm3}
\bibinfo{author}{\bibfnamefont{M.}~\bibnamefont{Bartenstein}},
  \bibinfo{author}{\bibfnamefont{A.}~\bibnamefont{Altmeyer}},
  \bibinfo{author}{\bibfnamefont{S.}~\bibnamefont{Riedl}},
  \bibinfo{author}{\bibfnamefont{S.}~\bibnamefont{Jochim}},
  \bibinfo{author}{\bibfnamefont{C.}~\bibnamefont{Chin}},
  \bibinfo{author}{\bibfnamefont{J.~H.} \bibnamefont{Denschlag}},
  \bibnamefont{and} \bibinfo{author}{\bibfnamefont{R.}~\bibnamefont{Grimm}},
  \bibinfo{journal}{Phys. Rev. Lett.} \textbf{\bibinfo{volume}{92}},
  \bibinfo{pages}{203201} (\bibinfo{year}{2004}{\natexlab{a}}).

\bibitem[{\citenamefont{Kinast et~al.}(2005)\citenamefont{Kinast, Turlapov,
  Thomas, Chen, Stajic, and Levin}}]{ThermoScience}
\bibinfo{author}{\bibfnamefont{J.}~\bibnamefont{Kinast}},
  \bibinfo{author}{\bibfnamefont{A.}~\bibnamefont{Turlapov}},
  \bibinfo{author}{\bibfnamefont{J.~E.} \bibnamefont{Thomas}},
  \bibinfo{author}{\bibfnamefont{Q.~J.} \bibnamefont{Chen}},
  \bibinfo{author}{\bibfnamefont{J.}~\bibnamefont{Stajic}}, \bibnamefont{and}
  \bibinfo{author}{\bibfnamefont{K.}~\bibnamefont{Levin}},
  \bibinfo{journal}{Science} \textbf{\bibinfo{volume}{307}},
  \bibinfo{pages}{1296} (\bibinfo{year}{2005}), \bibinfo{note}{published online
  27 January 2005; doi:10.1126/science.1109220}.

\bibitem[{\citenamefont{Chen et~al.}(2005{\natexlab{a}})\citenamefont{Chen,
  Stajic, Tan, and Levin}}]{ourreview}
\bibinfo{author}{\bibfnamefont{Q.~J.} \bibnamefont{Chen}},
  \bibinfo{author}{\bibfnamefont{J.}~\bibnamefont{Stajic}},
  \bibinfo{author}{\bibfnamefont{S.~N.} \bibnamefont{Tan}}, \bibnamefont{and}
  \bibinfo{author}{\bibfnamefont{K.}~\bibnamefont{Levin}},
  \bibinfo{journal}{Phys. Rep.} \textbf{\bibinfo{volume}{412}},
  \bibinfo{pages}{1} (\bibinfo{year}{2005}{\natexlab{a}}).

\bibitem[{\citenamefont{Chen et~al.}(2006)\citenamefont{Chen, Stajic, and
  Levin}}]{ReviewJLTP}
\bibinfo{author}{\bibfnamefont{Q.~J.} \bibnamefont{Chen}},
  \bibinfo{author}{\bibfnamefont{J.}~\bibnamefont{Stajic}}, \bibnamefont{and}
  \bibinfo{author}{\bibfnamefont{K.}~\bibnamefont{Levin}},
  \bibinfo{journal}{Fizika Nizkikh Temperatur (Low Temp. Phys.)}
  \textbf{\bibinfo{volume}{32}}, \bibinfo{pages}{538} (\bibinfo{year}{2006}).

\bibitem[{\citenamefont{Bourdel et~al.}(2003)\citenamefont{Bourdel, Cubizolles,
  Khaykovich, Magalhaes, Kokkelmans, Shlyapnikov, and Salomon}}]{Salomon}
\bibinfo{author}{\bibfnamefont{T.}~\bibnamefont{Bourdel}},
  \bibinfo{author}{\bibfnamefont{J.}~\bibnamefont{Cubizolles}},
  \bibinfo{author}{\bibfnamefont{L.}~\bibnamefont{Khaykovich}},
  \bibinfo{author}{\bibfnamefont{K.~M.~F.} \bibnamefont{Magalhaes}},
  \bibinfo{author}{\bibfnamefont{S.~J. J. M.~F.} \bibnamefont{Kokkelmans}},
  \bibinfo{author}{\bibfnamefont{G.~V.} \bibnamefont{Shlyapnikov}},
  \bibnamefont{and} \bibinfo{author}{\bibfnamefont{C.}~\bibnamefont{Salomon}},
  \bibinfo{journal}{Phys. Rev. Lett} \textbf{\bibinfo{volume}{91}},
  \bibinfo{pages}{020402} (\bibinfo{year}{2003}).

\bibitem[{\citenamefont{Bartenstein
  et~al.}(2004{\natexlab{b}})\citenamefont{Bartenstein, Altmeyer, Riedl,
  Jochim, Chin, Denschlag, and Grimm}}]{Grimm2}
\bibinfo{author}{\bibfnamefont{M.}~\bibnamefont{Bartenstein}},
  \bibinfo{author}{\bibfnamefont{A.}~\bibnamefont{Altmeyer}},
  \bibinfo{author}{\bibfnamefont{S.}~\bibnamefont{Riedl}},
  \bibinfo{author}{\bibfnamefont{S.}~\bibnamefont{Jochim}},
  \bibinfo{author}{\bibfnamefont{C.}~\bibnamefont{Chin}},
  \bibinfo{author}{\bibfnamefont{J.~H.} \bibnamefont{Denschlag}},
  \bibnamefont{and} \bibinfo{author}{\bibfnamefont{R.}~\bibnamefont{Grimm}},
  \bibinfo{journal}{Phys. Rev. Lett.} \textbf{\bibinfo{volume}{92}},
  \bibinfo{pages}{120401} (\bibinfo{year}{2004}{\natexlab{b}}).

\bibitem[{\citenamefont{Regal et~al.}(2005)\citenamefont{Regal, Greiner,
  Giorgini, Holland, and Jin}}]{Jin6}
\bibinfo{author}{\bibfnamefont{C.~A.} \bibnamefont{Regal}},
  \bibinfo{author}{\bibfnamefont{M.}~\bibnamefont{Greiner}},
  \bibinfo{author}{\bibfnamefont{S.}~\bibnamefont{Giorgini}},
  \bibinfo{author}{\bibfnamefont{M.}~\bibnamefont{Holland}}, \bibnamefont{and}
  \bibinfo{author}{\bibfnamefont{D.~S.} \bibnamefont{Jin}},
  \bibinfo{journal}{\prl} \textbf{\bibinfo{volume}{95}},
  \bibinfo{pages}{250404} (\bibinfo{year}{2005}).

\bibitem[{\citenamefont{Partridge et~al.}(2006)\citenamefont{Partridge, Li,
  Kamar, an~Liao, and Hulet}}]{Hulet5}
\bibinfo{author}{\bibfnamefont{G.~B.} \bibnamefont{Partridge}},
  \bibinfo{author}{\bibfnamefont{W.}~\bibnamefont{Li}},
  \bibinfo{author}{\bibfnamefont{R.~I.} \bibnamefont{Kamar}},
  \bibinfo{author}{\bibfnamefont{Y.}~\bibnamefont{an~Liao}}, \bibnamefont{and}
  \bibinfo{author}{\bibfnamefont{R.~G.} \bibnamefont{Hulet}},
  \bibinfo{journal}{Science} \textbf{\bibinfo{volume}{311}},
  \bibinfo{pages}{503} (\bibinfo{year}{2006}).

\bibitem[{\citenamefont{Zwierlein et~al.}(2006)\citenamefont{Zwierlein,
  Schirotzek, Schunck, and Ketterle}}]{Ketterle6}
\bibinfo{author}{\bibfnamefont{M.~W.} \bibnamefont{Zwierlein}},
  \bibinfo{author}{\bibfnamefont{A.}~\bibnamefont{Schirotzek}},
  \bibinfo{author}{\bibfnamefont{C.~H.} \bibnamefont{Schunck}},
  \bibnamefont{and} \bibinfo{author}{\bibfnamefont{W.}~\bibnamefont{Ketterle}},
  \bibinfo{journal}{Science} \textbf{\bibinfo{volume}{311}},
  \bibinfo{pages}{5760} (\bibinfo{year}{2006}).

\bibitem[{\citenamefont{Viverit
  et~al.}(2004{\natexlab{a}})\citenamefont{Viverit, Giorgini, Pitaevskii, and
  Stringari}}]{Viverit}
\bibinfo{author}{\bibfnamefont{L.}~\bibnamefont{Viverit}},
  \bibinfo{author}{\bibfnamefont{S.}~\bibnamefont{Giorgini}},
  \bibinfo{author}{\bibfnamefont{L.}~\bibnamefont{Pitaevskii}},
  \bibnamefont{and}
  \bibinfo{author}{\bibfnamefont{S.}~\bibnamefont{Stringari}},
  \bibinfo{journal}{Phys. Rev. A} \textbf{\bibinfo{volume}{69}},
  \bibinfo{pages}{013607} (\bibinfo{year}{2004}{\natexlab{a}}).

\bibitem[{\citenamefont{Regal and Jin}(2003)}]{Jin2}
\bibinfo{author}{\bibfnamefont{C.~A.} \bibnamefont{Regal}} \bibnamefont{and}
  \bibinfo{author}{\bibfnamefont{D.~S.} \bibnamefont{Jin}},
  \bibinfo{journal}{Phys. Rev. Lett.} \textbf{\bibinfo{volume}{90}},
  \bibinfo{pages}{230404} (\bibinfo{year}{2003}).

\bibitem[{\citenamefont{DeMarco and Jin}(1999)}]{Jin}
\bibinfo{author}{\bibfnamefont{B.}~\bibnamefont{DeMarco}} \bibnamefont{and}
  \bibinfo{author}{\bibfnamefont{D.~S.} \bibnamefont{Jin}},
  \bibinfo{journal}{Science} \textbf{\bibinfo{volume}{285}},
  \bibinfo{pages}{1703} (\bibinfo{year}{1999}).

\bibitem[{\citenamefont{Chen et~al.}(2005{\natexlab{b}})\citenamefont{Chen,
  Stajic, and Levin}}]{ChenThermo}
\bibinfo{author}{\bibfnamefont{Q.~J.} \bibnamefont{Chen}},
  \bibinfo{author}{\bibfnamefont{J.}~\bibnamefont{Stajic}}, \bibnamefont{and}
  \bibinfo{author}{\bibfnamefont{K.}~\bibnamefont{Levin}},
  \bibinfo{journal}{\prl} \textbf{\bibinfo{volume}{95}},
  \bibinfo{pages}{260405} (\bibinfo{year}{2005}{\natexlab{b}}).

\bibitem[{\citenamefont{Stajic et~al.}(2005)\citenamefont{Stajic, Chen, and
  Levin}}]{JS5}
\bibinfo{author}{\bibfnamefont{J.}~\bibnamefont{Stajic}},
  \bibinfo{author}{\bibfnamefont{Q.~J.} \bibnamefont{Chen}}, \bibnamefont{and}
  \bibinfo{author}{\bibfnamefont{K.}~\bibnamefont{Levin}},
  \bibinfo{journal}{Phys. Rev. Lett.} \textbf{\bibinfo{volume}{94}},
  \bibinfo{pages}{060401} (\bibinfo{year}{2005}).

\bibitem[{\citenamefont{Leggett}(1980)}]{Leggett}
\bibinfo{author}{\bibfnamefont{A.~J.} \bibnamefont{Leggett}}, in
  \emph{\bibinfo{booktitle}{Modern Trends in the Theory of Condensed Matter}}
  (\bibinfo{publisher}{Springer-Verlag}, \bibinfo{address}{Berlin},
  \bibinfo{year}{1980}), pp. \bibinfo{pages}{13--27}.

\bibitem[{\citenamefont{Viverit
  et~al.}(2004{\natexlab{b}})\citenamefont{Viverit, Giorgini, Pitaevskii, and
  Stringari}}]{Stringari}
\bibinfo{author}{\bibfnamefont{L.}~\bibnamefont{Viverit}},
  \bibinfo{author}{\bibfnamefont{S.}~\bibnamefont{Giorgini}},
  \bibinfo{author}{\bibfnamefont{L.~P.} \bibnamefont{Pitaevskii}},
  \bibnamefont{and}
  \bibinfo{author}{\bibfnamefont{S.}~\bibnamefont{Stringari}},
  \bibinfo{journal}{Phys. Rev. A} \textbf{\bibinfo{volume}{69}},
  \bibinfo{pages}{013607} (\bibinfo{year}{2004}{\natexlab{b}}).

\bibitem[{\citenamefont{Chiofalo et~al.}(2005)\citenamefont{Chiofalo, Giorgini,
  and Holland}}]{Chiofalo2005a}
\bibinfo{author}{\bibfnamefont{M.}~\bibnamefont{Chiofalo}},
  \bibinfo{author}{\bibfnamefont{S.}~\bibnamefont{Giorgini}}, \bibnamefont{and}
  \bibinfo{author}{\bibfnamefont{M.}~\bibnamefont{Holland}},
  \bibinfo{journal}{cond-mat/0512460}  (\bibinfo{year}{2005}).

\end{thebibliography}

\end{document}